Shteryo Nozharov[1]

# THE INSTITUTIONAL ECONOMICS OF COLLECTIVE WASTE RECOVERY SYSTEMS: AN EMPIRICAL INVESTIGATION

*The main purpose of the study is to develop the model for transaction costs measurement in the Collective Waste Recovery Systems. The methodology of New Institutional Economics is used in the research. The impact of the study is related both to the enlargement of the limits of the theory about the interaction between transaction costs and social costs and to the identification of institutional failures of the EU concept for circular economy. A new model for social costs measurement is developed.*
*Keywords: circular economy, transaction costs, extended producer responsibility*
*JEL: A13, C51, D23, L22, Q53*

## 1. Introduction

The Institutional Economics has come a long way since the renowned work of Hamilton (1919). For the last 100 years it was developed and the focus was put over various economic fields. The following publications appeared: The Institutional Economics of Foreign Aid (Lensink, 2003), The Institutional Economics of Corruption and Reform (Spechler, 2009), The institutional economics of biodiversity, biological materials, and bioprospecting (Polski, 2005).

In the current publication it is made an attempt to be developed institutional basis in the economic analysis of a specific economic sector – Collective Waste Recovery Systems. The concept for *circular economy* provokes increased scientific interest. The principles of the circular economy are being studied (Geissdoerfer, Savaget, Bocken and Hultink, 2017). The definition for circular economy (Kirchherr, Reike and Hekkert, 2017) and its content (Yuan, Bi and Moriguichi, 2008) have been clarified. However, the aforementioned publications used mainly the methodology of Ecological Economics. There is lack of in-depth analysis based on the methodology of the New Institutional Economics.

One of the main purposes of the current research is to analyze the possibilities, social costs to be deducted only by the basis of transaction costs. Such kind of hypothesis has not been proven yet because it is perceived as an impossible one (Berger, 2017).

---

[1] *Shteryo Nozharov,Ph.D., Full-time faculty member, Department of Economics, University of National and World Economy, Bulgaria, e-mail:* [nozharov@unwe.bg](nozharov@unwe.bg)

Another goal of the study is to reveal the institutional failures of the EU concept for circular economy (COM/2015/0595). For that purpose, the model for transaction costs measurement in Collective Waste Recovery Systems (Nozharov, 2018) will be used.

## 2. Theoretical background

### 2.1. Institutional failure

The main institutional components of the current study are the company (private hierarchy) and the government (public hierarchy). The market is examined as a set of rules and prices which determine the coincidence between supply and demand needs for the whole quantity of goods. The following risk factors for the market equilibrium are identified: wide variety of traded goods, possible delays in supply, business investment distortions, discrepancies between household needs and purchasing power, government failures, oligopolies and monopolies, externalities, information asymmetry and etc., (Pitelis, 1992).

One of the opportunities such risks to be overcome according to the economic theory is the free markets to be replaced by hierarchies (Kroszner and Putterman, 2009). The main purpose is this fact to cause more efficient resource allocation and to solve the problems related to the market failures. For example, the government as a "public hierarchy" could intervene in the market and compensate the market failures (Arrow 1970, Coase 1960, Pitelis 1992). The government could exercise its power to overcome the market failures with lower costs than those of the market and the "private hierarchies" (companies). The government however could not replace the private sector entirely because it has problems with its effectiveness (Lewin 1982, Mueller 2003). The functioning of hierarchies is related to low power stimuli in comparison to the markets and this fact limits their effectiveness (Williamson 1973, Valentinov and Chatalova 2014).

The EU established a specific type of hierarchy in the field of waste management in order to be overcome the existing market ineffectiveness. These hierarchies exist in the current European legislation and they are called "Collective financing schemes to meet obligations arising from the extended producer responsibility"(Directive 2012/19/EU). The new European legislation for "circular economy" make provisions for such schemes. In this legislation the schemes are called "organisations implementing extended producer responsibility" on behalf of the obligated producers (COM/2015/0595). The term "collective waste recovery systems" which is used in the current publication is equal to the meaning of the aforementioned schemes.

These collective waste recovery systems are a specific type of hierarchy, which is a combination of private and public hierarchies. Obviously the main purpose of EU when introducing such schemes in the legislation for circular economy is to overcome the disadvantages of the private and public hierarchies. For example, if a company meets individually its obligations for extended producer responsibility (Directive 2008/98/EC), it will not have enough financial capacity and its transaction costs will be high (Dubois, 2012). On the other side, if the government meets the obligations for extended producer responsibility on behalf of the companies, this activity will be much more inefficient (Mueller, 2003). However, there is a risk in the EU legislation for circular economy to be combined not the advantages of the private and public hierarchies but their disadvantages. One of the main purposes of the current publication is the aforementioned issue to be clarified.

2.2.     *Relationship between social and transaction costs*

There are various definitions for the term "Social cost". The mainstream definition assumes that they are a combination of the "private costs" of the transaction and the "externalities" imposed on the society as a result of the production and consumption of the traded good or service (Berger 2017, Pigou 1954). The current study assumes the wider definition of Kapp (1971). According to his definition social costs are the sum of the direct and indirect losses, endured by third party or by the society as a result of the unrestrained economic activity. Lewin (1982) sees the possibility social costs to be widespread examined as missed social benefits from the viewpoint of the concept for "opportunity cost".

The relationship between social costs and transaction costs is studied by Coase (1960). According to him the presence of externalities is explained with the high level of transaction costs in the contract. The solution of the problem he sees in the predefinition of the property rights. Valentinov and Chatalova (2014) put the attention on the paradox that social costs are generated by the same hierarchies which successfully minimize the transaction costs.

Social costs in the field of environmental policy are defined by the U.S. Environmental Protection Agency (EPA-USA, 2000a). According to EPA, "total social cost" are the sum of the opportunity costs made by the society as a result of the pursued regulation policy. This is the value of goods and services lost by society as a result of the use of resources in compliance with the requirements of the regulation and the reduction of the final output. The external benefits for the society (for health and environment) which compensate the externalities are not taken into account. Total social cost in the EPA model consist of five components: *Real-resource compliance costs, Government regulatory costs, Social welfare losses, Transitional costs, Indirect costs*. They are measured by the analysis of the demand and supply elasticity of the goods and services. Transaction costs are included in the *Transitional costs,* together with level of unemployment, companies' bankruptcy and etc. They are presented as a small part in one of the five components of the social costs. According to EPA transaction costs arose as a result of the implementation of new stimuli based policies such as a tradable permits program.

All of the examined publications in the field of social costs in the waste management, use the model of EPA-USA (2000a). Kinnaman, Shinkuma and Yamamoto, (2014) studied the social optimal level of recycling in Japan. What has been achieved by their model is the fact that not only external costs are taken into account but also the external benefits of recycling. Jamasb and Nepal, (2010) also use the model of EPA in their publication. They carry out a social cost–benefit analysis of waste-to-energy in the UK. They admit that transaction costs are a small part of the social costs in terms of implementation of policies based on new economic tools. The publication of Dijkgraaf and Vollebergh is in the same sense (2004). The one and only publication that uses different approach to measure the social costs is that of Jones, (2010). In Jone's publication social costs are examined through the focus of social capital theory. According to this theory social costs have no economic value and they are an indicator for environmental behavior. Social costs consist of four factors: social trust, institutional trust, compliance with social norms and participation in social networks.

The purpose of the current publication is to prove that social costs could be deducted only by the transaction costs. This hypothesis goes beyond the mainstream economic theory for social costs, examined in the paragraph above.

## 3. Study area, methods and results

The economic model for "extended producer responsibility" and collective waste management systems in the EU is the scope of the research. There will be used statistical data for Bulgaria since 2007 as a EU member state in order the model to be empirically tested. The statistical period of research is over ten years, which allows the summarized conclusions about the main hypothesis of the research (the possibility social costs to be deducted by the transaction costs and the presence of institutional failures in the EU concept for circular economy) to be valid. There will be analyzed the waste oils flow (Regulation (EC) No 2150/2002, Annex III - Waste statistical nomenclature, "*01.3 - Used oils*") because of its economic significance and the fact this issue is not examined in depth by the researchers. Quantitative and qualitative analyses will be applied.

### 3.1. *Qualitative methods*

The main purpose of the qualitative analysis is to analyze the quality and impact of the institutional environment over the collective waste recovery systems. The financial analysis methodology based on four indicators will be used. The Commercial register (2018) will be used as a main source of statistical information. According to the data the average number of collective waste recovery systems in the field of waste oils management for the period 2007-2017 in Bulgaria is 5. The average number of companies in the sector of hazardous waste management in Bulgaria is 50. The financial analysis includes the following four indicators: profit, value of the long-term tangible assets, average number of the persons employed and total liquidity ratio. The results of the analysis are presented in table 1.

*Table 1. Financial analysis of the collective waste recovery systems in Bulgaria in the field of waste oils management based on four indicators (2007-2017)*

| № | Indicator | Amount |
|---|---|---|
| 1.1 | Average profit for the period - 5 collective waste recovery systems | —11.483 |
| | (annually) | euro |
| 1.2 | Average profit for the period – 50 companies (entire sector) for hazardous waste management (annually) | 29.950 euro |
| 1.3 | Share of the collective waste recovery systems in the entire sector for hazardous waste management, which work at a loss | 60% |
| 1.4 | Highest average profit for the period, made by a collective waste recovery system (annually) | 3.125 euro |
| 2.1 | Average annual value of the long-term tangible assets - 5 collective waste recovery systems (annually) | 3.245 euro |
| 2.2 | Average annual value of the long-term tangible assets for the entire sector - 50 companies in the field of hazardous waste management (annually) | 1.143.500 euro |
| 3.1 | Average number of persons employed in collective waste recovery systems (annually) | 2 persons |
| 3.2 | Average number of persons employed in the entire sector - 50 companies in the field of hazardous waste management (annually) | 95 persons |
| 4.1 | Average total liquidity ratio of a collective waste recovery system | 0.8491 |
| 4.2 | Average total liquidity ratio for the entire sector – 50 companies in the field of hazardous waste management | 1.1701 |

*Source: Author's calculations*

The results of the financial analysis based on the four indicators show that collective waste recovery systems in Bulgaria in the field of waste oils management for the period 2007-2017 are decapitalized, unprovided with employees and long-term tangible assets and they have no liquidity. They work at loss, while the other companies in the sector work at profit. The average annual value of the long-term assets is approximately 3 000 euro which amount is equal to the amount of the average monthly wages in the EU. This amount is 352 times lower than the average value of the long-term assets in the entire sector of companies in the field of hazardous waste management in Bulgaria for the period 2007-2017. The average annual personnel in the collective waste recovery systems consists of 2 employees which number could be compared to the personnel of a bookstore. This number is 47 times lower than the average number of persons employed in the entire sector of companies in the field of waste oils management in Bulgaria. The total liquidity ratio of collective waste recovery systems is under 1 (0.84), while the total liquidity ratio for the entire sector is over 1 (1.17). There is a risk for collective waste management systems of servicing their depts.

Consequently, the following conclusion could be done – there are institutional failures for the execution of the extended producer responsibility by collective waste recovery systems. Bulgaria is a member of the EU since 2007 and more than ten years the country enforce the EU legislation. Since all collective waste recovery systems, involved in the management of the same type of waste oils suffer from the same financial defects, the problem is obviously in the institutional environment. This leads to an institutional failure.

*3.2.    Quantitative methods*

The purpose of the quantitative analysis is to examine the level of the social costs as a result of the institutional failure of the collective waste recovery systems. The analysis is done in the example of Bulgaria, but it could be implemented in every EU member-state. The existence of such type of costs was identified in the qualitative analysis. In order the social costs to be measured, there will be used the model for transaction costs measurement in Collective Waste Recovery Systems from the publication of Nozharov (2018). The first level of the model measures the so called "Social Public Costs":

**Social Public Costs (SPC)=** *the quantity of regenerated waste oil with collective systems – the quantity of regenerated waste oil without collective systems*    (1)

These costs are called Social Public Costs because they are differently measured compared to the conventional concept for "Social cost", according to which social costs are the sum of Private Costs and External Costs (Berger 2017, Pigou 1954). In the current publication the model for social costs measurement will be developed as follows:

**SPC=** *the quantity of regenerated waste oil with collective systems – the capacity of technology in the private processing companies in the relevant country in accordance with Directive 2010/75/EU*    (1.1)

or

**SPC=** *the quantity of regenerated waste oil with collective systems – consumer quantity demanded of produced goods from regenerated waste oil according to the available statistical data*    (1.2)

Due to the limited number of pages of the current paper, there will be presented only the results of the calculations. In equation 1.1, the capacity of technology in the private processing companies in Bulgaria is *two times* over the quantity of the regenerated waste oil which is reported as a result of the Collective Waste Recovery Systems activity. The statistical data about the capacity of technology of the private processing companies are taken from the registers of the Bulgarian Ministry of Environment in accordance with Directive 2010/75/EU. The statistical data, concerning the quantity of regenerated waste oil with collective systems are also taken from the registers of the Ministry of Environment of Bulgaria in accordance with Directive 2008/98/EC and Regulation (EC) No 2150/2002.

In equation 1.2, the consumer quantity demanded of produced goods from regenerated waste oil exceeds *seven times* the quantity of the regenerated waste oil which amount is reported as a result of the Collective Waste Recovery Systems activity. There must be taken into account also the quantity of the imported goods of regenerated waste oils. Even in this case, the consumer demanded quantity of goods, produced from regenerated waste oil will be two times higher than the quantity of the regenerated oil. On the other hand, there must be taken into account the amount of the exported goods, produced from regenerated waste oil, which amount could compensate the amount of the imported goods.

This means that the existence of Collective Waste Recovery Systems is not necessary when the processing companies could buy the waste oil from companies with extended producer responsibility.

Consequently, the social public costs will equal the total costs of the 5 Collective Waste Recovery Systems in Bulgaria. These social public costs could be measured by the opportunity costs, paid by the companies for the services provided by the Collective Waste Recovery Systems. The direct opportunity costs will be measured as loss for the society as a result of the missed opportunity these companies to invest money in new green technologies.

**Conclusions**

In accordance with the mainstream economic theory, social costs equal *private costs + external costs* (Berger 2017, Pigou 1954). However, the current publication disputes the conventional definition for social costs. First of all, for Collective Waste Recovery Systems we cannot talk about "Private Costs" in accordance with the classical understanding of the term. This due to the fact that the main purpose of the Collective Waste Recovery Systems is not to produce goods which will be traded at the private markets, but to make high profit. The main purpose of such systems is to lower their total costs through the implication of the "extended producer responsibility" principle (COM/2015/0595). Secondly, for Collective Waste Recovery Systems we cannot talk about "External Costs". This due to the fact that the main purpose of Collective Waste Recovery Systems is to eliminate "External Costs", rather than to create them. Their mission is to clean the environment of waste materials, rather than to create them. In this way, their social costs cannot be calculated as a sum of the "Private Costs" and "External Costs". These costs could only be calculated as a function of the transaction costs in accordance with the model presented in the publication of Nozharov (2018). Consequently, the social costs of Collective Waste Recovery Systems will be a function of their transaction costs. The amount of these social costs will depend on the institutional ineffectiveness which equals the sum of the total costs of collective waste recovery systems. The loss for the society will be measured by the opportunity costs.

If the market structure of the waste recovery sector is an oligopoly, then Collective Waste Recovery Systems are inefficient and the amount of their transaction and social costs will be at the maximum.

Transaction costs in the field of environmental policy do not exist only in the implementation of incentive-based policies, as it is stated by the EPA-USA (2000). They also exist in the case of distortions of the institutional environment where the environmental policy is implemented. Because of the limited number of pages of the current paper, the proposals for developing the EU concept for circular economy will be presented in a separate research study.

## Acknowledgements

I would like to thank to Prof. Sebastian Berger, author of *The Social Costs of Neoliberalism - Essays on the Economics of K. William Kapp (2017),* for his critical remarks, concerning the idea of the current study.

I would like to thank also to Prof. Rumen Gechev, project manager of university scientific project NI-19/2018, UNWE-Sofia, for the financial support of the current study.